\documentclass[12pt]{article}
\usepackage{epsfig,amssymb}
\usepackage{graphicx,color}

\newcommand{\be}{\begin{equation}}
\newcommand{\ee}{\end{equation}}
\newcommand{\bear}{\begin{eqnarray}}
\newcommand{\eear}{\end{eqnarray}}
\newcommand{\ba}{\begin{array}}
\newcommand{\ea}{\end{array}}

\begin{document}

\begin{flushright}
{\tt hep-th/0611053}
\end{flushright}

\vspace{5mm}

\begin{center}
{{{\Large \bf  Evolution of Tachyon Kink with Electric Field}}\\[14mm]
{Inyong Cho}\\[2.5mm]
{\it Department of Physics and BK21 Physics Research Division, \\
Sungkyunkwan University, Suwon 440-746, Korea}\\
{\tt iycho@skku.edu}\\[5mm]
{O-Kab Kwon}\\[2.5mm]
{\it Center for Quantum Spacetime, \\
Sogang University, Seoul 121-742, Korea}\\
{\tt okabkwon@sogang.ac.kr}\\[4mm]
{Chong Oh Lee}\\[2.5mm]
{\it Department of Physics, \\
Chonbuk National University, Chonju 561-756, Korea}\\
{\tt cohlee@chonbuk.ac.kr} }
\end{center}

\vspace{10mm}

\begin{abstract}
We investigate the decay of an inhomogeneous D1-brane
wrapped on a $S^1$ with an electric field.
The model that we consider consists of an array of tachyon kink
and anti-kink with a constant electric flux.
Beginning with an initially static configuration,
we numerically evolve the tachyon field with some perturbations
under a fixed boundary condition at diametrically opposite points
on the circle $S^1$.
When the electric flux is smaller than the critical value,
the tachyon kink becomes unstable;
the tachyon field rolls down the potential,
and the lower dimensional D0- and $\bar {\rm D}0$-brane become thin,
which resembles the caustic formation known for
this type of the system in the literature.
For the supercritical values of the electric flux,
the tachyon kink remains stable.
\end{abstract}

\newpage

\setcounter{equation}{0}
\section{Introduction}

Spatially inhomogeneous rolling tachyon
on an unstable D-brane has been studied
in boundary conformal field
theory~(BCFT)~\cite{Sen:2002vv,Larsen:2002wc,Rey:2003zj}, boundary string
field theory~(BSFT)~\cite{Ishida:2003cj}, and Dirac-Born-Infeld~(DBI)-type
effective field theory~\cite{Felder:2002sv,Mukohyama:2002vq,Berkooz:2002je,
Cline:2003vc,Felder:2004xu,Barnaby:2004dp,Barnaby:2004nk,
Panigrahi:2004qr,Saremi:2004yd,Canfora:2005mt,Kwon:2006wh}.
In BCFT, this subject was considered in the presence of spacetime-dependent
marginally deformed tachyon vertex operator at the worldsheet boundary.
The evolution of the resulting energy-momentum tensor is qualitatively
different from that of the homogeneous rolling
tachyon~\cite{Sen:2002nu,Sen:2002in}.
The energy density  evolves into a localized delta-function array
within a finite critical time.
These delta-function singularities were interpreted as codimension-one
D-branes~\cite{Sen:2002vv,Larsen:2002wc,Rey:2003zj}.

In lower-energy effective field theory approaches for unstable D-branes,
most of the actions involve only the first
derivatives of fields on the brane. Since the actions are represented by
truncating all the higher derivatives of fields, they are considered reliable
only when the second and higher derivatives are small.
Nonetheless, the effective
actions have been shown to reproduce various nontrivial aspects of the
unstable D-brane dynamics in special settings, for example,
the spatially homogeneous rolling tachyon~\cite{Sen:2002an,Sugimoto:2002fp,
Gibbons:2002tv,Kim:2003he} and lower dimensional D-branes as
static worldvolume solitons~\cite{Sen:2003tm,Lambert:2003zr,Brax:2003rs,
Kim:2003in,Afonso:2006ws,Kim:2006mg,Kim:2005tw}.

In an inhomogeneous time evolution of an unstable D-brane, however,
its dynamics is not governed by a truncated effective
action after a certain critical time.
Felder et. al. showed that caustics develop in some regions
during the evolution of the inhomogeneous tachyon field
with a runaway potential in the DBI-type action~\cite{Felder:2002sv}.
They treated the tachyon as a collisionless fluid and found
that the tachyon field becomes multi-valued in those regions
within a finite critical time.
They interpreted the caustics as a signal that the higher derivatives of
the tachyon field blow up~\cite{Felder:2002sv,Barnaby:2004nk}.
More examples for the pure tachyon evolution
have been investigated in Refs.~\cite{Mukohyama:2002vq,Cline:2003vc}.
They observed that the slope of the tachyon field near the kink
diverges within a finite time.
Authors identified this phenomenon with the caustic formation
observed in Ref.~\cite{Felder:2002sv}.

Interesting aspects of the low energy effective
field theory with an electric flux turned on, have been considered
in the tachyon vacuum~\cite{Gibbons:2000hf,Gibbons:2002tv,Kim:2003he,
Kwon:2003qn,Yee:2004ec,Kwon:2006wh}.
In this work, we shall examine the role of the electromagnetic fields
on an unstable D-brane
where those fields are varying in time and in space
with a nonzero flux.
The electric flux density here is
identified with the string charge density
on the D-brane~\cite{Yi:1999hd,Bergman:2000xf}.
When the tachyon and the electromagnetic fields
depend on time (or one spatial) coordinate only,
the equations of motion imply that all the electromagnetic fields
are constants~\cite{Kim:2003he,Kim:2003in}.
Consequently, the electromagnetic fields become simply parameters
of the tachyon solution.
However, in an inhomogeneous time evolution process,
the electromagnetic fields depend on both time and space
in general~\cite{Kwon:2006wh}.

In this paper, we shall consider a simple case in the DBI-type action
which contains a tachyon and a single electric field.
The electric field is turned on along the inhomogeneous direction
which we restrict to be a circle $S^1$.
Assuming that the tachyon and the electric field
depend on time and one spatial coordinate,
we numerically solve the master equation
of the tachyon field under a fixed boundary condition at
diametrically opposite points, which is closely related to the formation of
the ${\rm D}0$-$\bar {\rm D}0$ pair located at diametrically
opposite points on the circle $S^1$~\cite{Sen:1998tt,Sen:1998ex,Sen:2003zf}.

We find that there exists a critical value of the string charge density
under which the kink is unstable.
The kink approaches a similar state
which resembles the caustic formation
observed in the pure tachyon case~\cite{Cline:2003vc}.
If the string charge density is increased above the critical value,
we find that the tachyon kink becomes stable.

In Section 2, we present the D$p$-brane model
with a tachyon and an electric field in the DBI-type action.
In Section 3, we numerically solve the field equation
and discuss the stability.
In Section 4, we analyze the stability in a semi-analytic manner,
and conclude in Section 5.

\section{D$p$-brane Model with Tachyon and Electric Field}

We shall consider the dynamics of the tachyon field
described by DBI-type action
in the presence of gauge field interactions.
With general gauge field interactions,
the tachyon and the gauge fields are coupled
in a very complicated manner.
Therefore, we shall consider the simplest case
in which the tachyon field depends
on time and on one spatial coordinate only
while the electric field is turned on
along that inhomogeneous spatial direction.

The DBI-type action with gauge field interactions on an unstable
D$p$-brane in flat space
is given by~\cite{Garousi:2000tr,Bergshoeff:2000dq,Kluson:2000iy}
\bear\label{AC}
&&\hskip 2.5cm S = - {\cal T}_{p} \int d^{p+1}x\, V(T) \sqrt{-X},
\\
&&X= \det(\eta_{\mu\nu} + \partial_\mu T\partial_\nu T + F_{\mu\nu}),
\quad (\mu,\,\nu = 0,1,\dots,p),
\label{X}
\eear
where
${\cal T}_p$ is the unstable D$p$-brane tension,
$V(T)$ is a runaway tachyon potential,
and $F_{\mu\nu}$ is the field-strength tensor
for the gauge field $A_\mu$.
The equations of motion for the tachyon and the gauge field read
\bear
&&\partial_{\mu}\left(\frac{V C^{\mu\nu}_{{\rm S}}}{\sqrt{-X}}
\partial_{\nu}T\right)
+ \sqrt{-X}\frac{dV}{dT}=0,
\label{te1} \\
&& \partial_{\mu}\left(\frac{V C^{\mu\nu}_{{\rm A}}}{\sqrt{-X}}\right)=0,
\label{ge1}
\eear
where $C^{\mu\nu}_{{\rm S(A)}}$ is the symmetric (anti-symmetric) part of
the cofactor for the matrix $(\eta + \partial T\partial T + F)_{\mu\nu}$.
We consider the tachyon and the electric field along the inhomogeneous
direction living in the worldvolume of the D$p$-brane,
\bear\label{TE}
T=T(x^0,x^1), \quad E\equiv F_{01}(x^0, x^1),
\eear
and, for simplicity, turn off all the other components
of the electromagnetic field.
For the unstable D1-brane, the above field ans\"atze~(\ref{TE})
are the general setting with no restriction.
With these fields on the unstable D$p$-brane,
the quantity $X$ and the cofactor matrix are simplified as
\bear\label{XX}
X &=& -(1-\dot T^2 + T^{'2}-E^2),\\
\left( C^{\mu\nu} \right) &=& \left(\begin{array}{ccc}
1 +   {T^{'}}^2 & E -  {\dot T}T^{'} \\
-E -  {\dot T}T^{'} & -1 +  {\dot T}^2 \end{array} \right),
\label{XC}
\eear
where $\dot T\equiv \partial_0 T$, $T' \equiv \partial_1 T$ and all
other components of $C^{\mu\nu}$ are trivial, i.e.,
$C^{\mu\nu}=\delta^{\mu\nu}$.

The solution to the gauge field equation~(\ref{ge1}) is then simply
\bear\label{ge2}
\Pi\equiv \beta E = \mbox{constant},
\eear
where $\Pi \equiv \partial {\cal L}/\partial  (\partial_0A_1)
=\partial {\cal L}/\partial E$
is the conjugate momentum for the gauge field $A_1$
in the Weyl gauge $A_0=0$.
We also have a defined quantity
\bear
\beta \equiv \frac{{\cal T}_p V}{\sqrt{-X}}.
\eear
Using Eq.~(\ref{ge2}), we can express the electric field $E$
with the tachyon field $T$ and the constant electric flux $\Pi$ as
\bear\label{E1}
E^2 = \frac{\Pi^2(1-\dot T^2 + {T^{'}}^2)}{\Pi^2 +
{\cal T}_p^2 V^2}.
\eear
The tachyon field equation~(\ref{te1})
which completely governs the dynamics of the unstable D-brane,
is now written as
\bear\label{te2}
(1 + {T'}^2)\ddot T - (1 - \dot T^2)T'' - 2 \dot T T'\dot T'
+ \frac{{\cal T}_p^2 V(1 -\dot T^2 + {T'}^2)}{\Pi^2
+ {\cal T}^2_p V^2} \frac{dV}{dT} =0.
\eear
Setting $\Pi=0$ in the above equation~(\ref{te2}),
we obtain the field equation for
the pure tachyon~\cite{Felder:2002sv}.

The components of the energy-momentum tensor are
\bear
\label{T00}
T^{00} &=& \beta (1+{T'}^2)
\\
\label{T01}
T^{01} &=& -\beta\dot{T}T',\\
\label{T11}
T^{11} &=& -\beta(1-\dot{T}^2).
\eear

Before we close this section, let us comment on what was studied
for the pure tachyon case ($\Pi=0$) in a relation to a caustic formation.
It is well-known that the tachyon potential for large $T$ exponentially
decays, $V(T) \sim e^{-\alpha T}$ with an arbitrary positive
constant $\alpha$~\cite{Sen:2002an}.
Then for the case of pure tachyon field ($\Pi = 0$), the tachyon
equation (\ref{te2}) is reduced to~\cite{Felder:2002sv}
\bear\label{te3}
(1 + {T'}^2)\ddot T - (1 - \dot T^2)T'' - 2 \dot T T'\dot T'
= \alpha \left(1-\dot T^2 + {T'}^2\right).
\eear
It was shown in the
literature~\cite{Felder:2002sv,Cline:2003vc,Felder:2004xu,Barnaby:2004nk}
that Eq.~(\ref{te3}) develops a caustic
at a finite time, which we shall reconsider in Sec. 4.3.

On the other hand, with a nonvanishing electric flux $\Pi$,
Eq.~(\ref{te2}) is rewritten as
\bear\label{te4}
(1 + {T'}^2)\ddot T - (1 - \dot T^2)T'' - 2 \dot T T'\dot T'
= -B(T) (1- \dot T^2 + {T'}^2),
\eear
where
\bear\label{BTT}
B(T) = \frac{{\cal T}_p^2 V}{ \Pi^2 + {\cal T}_p^2 V^2}\, \frac{dV}{dT}.
\eear
We shall discuss the role of the electric flux $\Pi$ for the 1/cosh-type
tachyon potential in Section 4 by investigating the behaviors of $B(T)$
for various $\Pi$'s.

\section{Evolution of Tachyon Kink Array on D1-brane}

In this section we investigate the time evolution of an
inhomogeneous tachyon kink array on an unstable D1-brane
in the presence of an electric field.
The D1-brane now contains a fundamental string with
a string charge density $\Pi$.
We shall also discuss the subsequent dynamical formation
of the lower dimensional D0/$\bar{\rm D}0$-brane.
Although we focus on the D1-brane for simplicity,
the results are generally applied to D$p$-branes
when only one spatial direction of them is inhomogeneous.
We solve the time-dependent field equation (\ref{te2}) numerically
with a well-known tachyon potential
\bear\label{tp}
V(T) = \frac{1}{\cosh\left(\frac{T}{\sqrt{2}}\right)}.
\eear

For the constant electric field, $E=E_0$, with the above potential
the static regular kink solution was obtained in Ref.~\cite{Kim:2003in}.
For $0< E_0^2 <1$, the solution reads
\bear\label{staticT}
T_0(x) = \sqrt{2} \sinh^{-1}\left[\sqrt{\frac{{\cal T}_1^2}{\beta_0^2-
\Pi^2} -1} \,\,
\sin\left(\frac{\sqrt{\beta_0^2-\Pi^2}}{\sqrt{2}\beta_0}\,x\right)\right],
\eear
where the constant $\beta_0 = \Pi/E_0$ is the value of $\beta$
for the static solution.
This solution represents a tachyon kink-anti-kink array
on a D1-brane, and reduces to a {\it pure} tachyon solution
when $\Pi = 0$~\cite{Lambert:2003zr}.

In order to investigate the tachyon evolution,
we evolve this static solution with a ``minimal" perturbation
by solving the time-dependent field equation~(\ref{te2}).
Here, the minimal perturbation is implied by
imposing the initial conditions,
\bear
\label{IC1}
T(t=0,x) &=& T_0(x),\\
\dot{T}(t=0,x) &=&  0.
\label{IC2}
\eear
In principle, if the the initial velocity of the tachyon field
is zero, the system never evolves;
it is easy to see from the field equation~(\ref{te2}) that
the static solution remains as the solution in the course of evolution.
However, in numerical simulations,
the intrinsic numerical errors coming from the
numerical scheme play a role of perturbation.
Once the numerical code is stable, the tiny numerical
error induces a minimal perturbation to the static system.
If the initial configuration is physically stable,
the system will not evolve
except exhibiting small fluctuations.
On the other hand,
if the system evolves from the initial configuration as time elapses,
it means that the system is physically unstable.
This is not from the numerical instability.
Therefore, applying such a minimal perturbation is very useful
for a system at present which is suspected
unstable to form a caustic at later times.

The boundary condition we impose is
\bear
\label{BC}
T(t,x_n) = 0,
\eear
where $x_n$ is the location of the $n$-th node
of the initial static configuration~(\ref{staticT}).
This boundary condition implies that we fix the spatial coordinates
of the kink/anti-kink centers during evolution.
By doing so, we may consider the time evolution of
only one patch (a half period) of the kink array,
and the rests are merely a duplication of that patch configuration.
Under the boundary condition (3.22), the compactification radius
is fixed by the periodicity of the static kink solution while it was
not fixed in the Sen's work in Ref.~\cite{Sen:1998ex}.

The initial conditions, (\ref{IC1}) and (\ref{IC2}),
imply the ``point symmetry" about
the kinks and the ``reflection symmetry" about the line $x=x_{max}$
where the tachyon field has the maximum value. The boundary
condition (\ref{BC}) that we impose preserves such symmetries in the course
of evolution. For the pure tachyon case, we can observe the
formation of the D0-$\bar {\rm D0}$ pair located at diametrically
opposite points on the circle ${\rm S}^1$ initiated by Sen,
and with the given boundary condition the ${\rm Z}_2$ symmetry is
preserved~\cite{Sen:1998tt,Sen:1998ex,Sen:2003zf}.
In the presence of electric flux,
we impose the boundary condition in the same way as we
do for the pure tachyon case, although the ${\rm Z}_2$
symmetry is now absent;
we believe that it is an appropriate and the simplest
set-up that we can consider
in the first step for the stability check.

\subsection{Pure Tachyon Case ($\Pi =0$)}

Let us first consider the pure tachyon case ($\Pi =0$)
in which caustic phenomena have been previously observed
in a bit different set-ups~\cite{Felder:2002sv,Cline:2003vc}.
In Ref.~\cite{Felder:2002sv}, the authors studied the caustic
problem with an arbitrary initial configuration,
and treated the tachyon field as collisionless fluid during the evolution.
They found that there develop some regions where
the tachyon field becomes multi-valued at a finite time.
In Ref.~\cite{Cline:2003vc},
the authors studied the tachyon evolution in a very similar way to ours,
but with different initial conditions and
a potential $V = \exp(-\alpha T^2)$.
Therefore, it is worthwhile to examine how the caustic phenomenon
arises in our set-up if it exists.

\begin{figure}
\centerline{\epsfig{figure=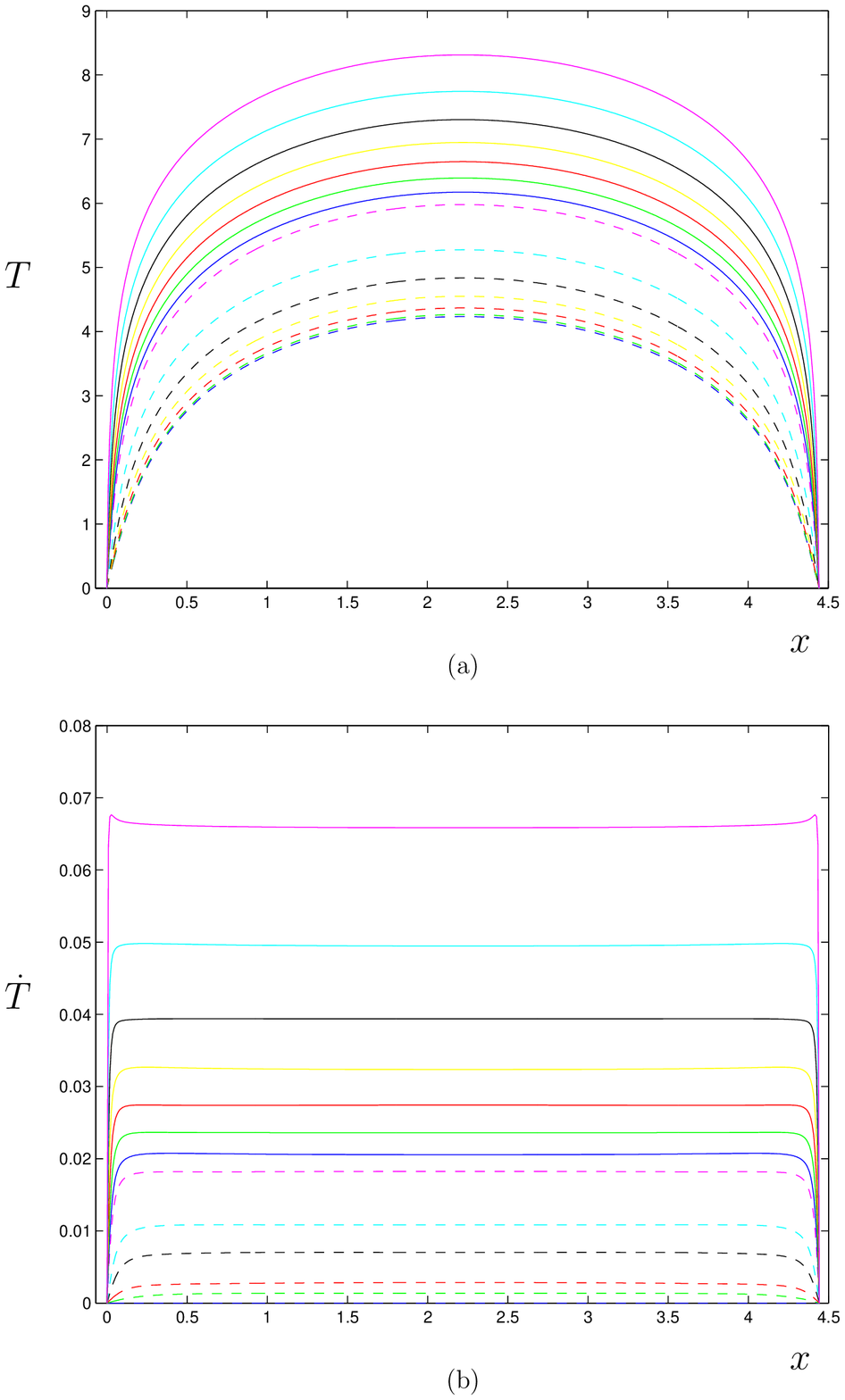,height=180mm}}
\caption{
Plot of (a) $T(t,x)$ and (b) $\dot{T}(t,x)$ for the pure tachyon case
with $\beta_0^2 = 10^{-2} {\cal T}_1^2$.
From the bottom, the lines correspond to
$t=0,50,100,...,300,310,320,...,370$.
(The line style and color are in the same order in the rest of figures.)
In (a), the tachyon field grows as it rolls down the run-away potential.
Near the kink, the slope increases rapidly as the tachyon fluid
flows to the kink.
In (b), $\dot{T}$ increases gradually in the bulk region
while it increases very rapidly near the kink.
If it is evolved further,
the slope gets very steep and a caustic is accompanied.
}
\label{FIG1}
\end{figure}

For $\Pi =0$, the static solution (\ref{staticT}) is simplified as
\bear\label{staticT0}
T_0(x) = \sqrt{2} \sinh^{-1}\left[\sqrt{\frac{{\cal T}_1^2}{\beta_0^2} -1} \,\,
\sin\left(\frac{x}{\sqrt{2}}\right)\right].
\eear
The proper range of $\beta_0$ is then given by
\bear
0<\beta_0^2  <{\cal T}_1^2.
\eear
In the limit of $\beta_0^2\to 0$,
the slopes of the tachyon profile (\ref{staticT0})
at the kink and the anti-kink sites become infinite,
and the region between the kink and
the anti-kink goes to the closed string vacuum.
However, the period of the static
tachyon configuration remains at the same constant value $2\sqrt{2}\pi$.
Therefore, the solution becomes an array of step functions with an infinite
height at each site of the kink and the anti-kink~\cite{Sen:2003tm,Kim:2003in}.
For various values of $\beta_0$ in the above range,
we have performed numerical calculations.
We observed that evolutions are more or less similar regardless of the
value of $\beta_0$.
We show numerical results in Fig.~\ref{FIG1} which exhibit a typical
evolution behavior.

Beginning with the static configuration and a minimal perturbation,
the tachyon field grows as it rolls down the potential.
The velocity of the tachyon field $\dot{T}$ gradually grows
everywhere except the fixed nodes where lower dimensional
D0/$\bar{\rm D}0$-branes are located.
The initial static configuration is never physically stable.
At late times, we can observe that the velocity near the kink
grows faster than in the bulk,
and the slope of the tachyon field gets very large.
This behavior is very similar to
what was observed in Ref.~\cite{Cline:2003vc},
so we identify it with the early stage of a caustic formation.
When it evolves further,
the numerical calculation diverges since the slope becomes too large
as in Ref.~\cite{Cline:2003vc}.

Due to the caustic behavior at late times,
it is not possible to observe the final decay products,
such as lower dimensional D-branes and tachyon matter,
in the inhomogeneous evolution of the unstable D-brane.
Although the main purpose of present work is
to investigate the stability of a given kink solution
using numerical techniques,
let us consider the decay products of our system
in an analytic manner.
Since we are considering the evolution of an unstable D-brane
on a compactified circle, the total energy of the system is finite.
The total energy of the initial kink (anti-kink) (\ref{staticT0})
is given by~\cite{Kim:2003in},
\bear
E_{{\rm kink}} = \int_{-\frac{\sqrt{2}}{2}\pi}^{\frac{\sqrt{2}}{2}\pi}
dx\, T^{00} = \sqrt{2}\pi {\cal T}_1 = {\cal T}_{0},
\eear
where the energy density $T^{00}$ is defined in Eq.~(\ref{T00}).
The energy of the initial static kink (anti-kink) gives
exactly the tension of D$0$ ($\bar {\rm D}0$)-brane
for the specific tachyon potential~(\ref{tp}).
Note that the energy does not depend on $\beta_0$.
In other words, the energy of the expected final decay product
D0 ($\bar {\rm D}0$) which corresponds to the kink (anti-kink)
in the thin limit ($\beta_0\to 0$) is exactly the same with that of
the kink (anti-kink) with an arbitrary $\beta_0$.
If the story is so, no tachyon matter is produced in principle.
For the other kink (anti-kink) solutions in more general runaway-type
tachyon potentials~\cite{Brax:2003rs},
the analytic approach becomes more complicated.

For our case, one can expect that the tachyon matter,
if it should be produced in the very final stage,
would come only from the minimal perturbation applied numerically,
but it is a negligible amount.
In the numerical evolution, however,
such an ideal final stage can never be reached
due to the caustic behavior as we described above,
so the final decay products are beyond description.

\subsection{Tachyon with Electric Flux ($\Pi \neq 0$)}

In this section, we consider the tachyon evolution
with an electric field turned on.
It is expected that the fundamental string which connects
the lower dimensional D0/$\bar{\rm D}0$-branes
modulates the tachyon flow and decay.

The electric flux $\Pi =\beta E$ remains constant
during evolution.
This is the solution to the gauge field equation
as was discussed in Sec. 2.
What is remaining is to solve the tachyon field
equation, which we shall do numerically with
applying $\Pi$ from a very small value.
For $\Pi \neq 0$,
the static solution (\ref{staticT}) is valid for
\bear\label{Pirange}
\beta_0^2 -{\cal T}_1^2 < \Pi^2 < \beta_0^2.
\eear
The value of $\beta_0$ is not restricted, in principle,
once $\Pi$ is ranged according to the above relation,
or vice versa.

\subsubsection{Unstable Solution}

The pure tachyon solution (\ref{staticT0}) is
a continuous limit ($\Pi \to 0$) of the static solution~(\ref{staticT})
with electric field.
Therefore, turning on tiny amount of electric flux
would not affect the evolution pattern so much,
and the result for small $\Pi$ will be
similar to that of the pure tachyon case.

\begin{figure}[!h]
\centerline{\epsfig{figure=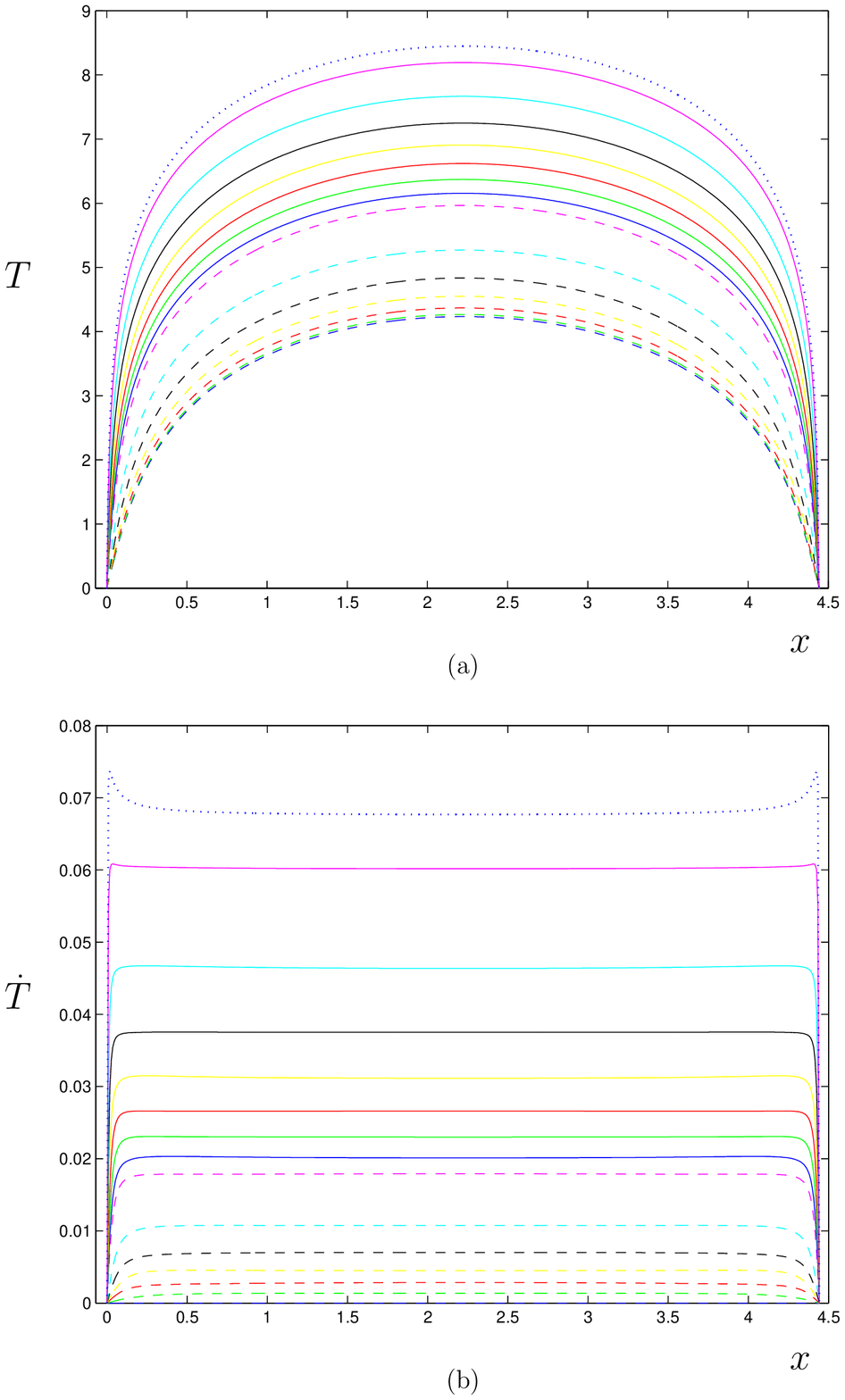,height=180mm}}
\caption{
Plot of (a) $T(t,x)$ and (b) $\dot{T}(t,x)$ for the tachyon-plus-electric
field case with $\beta_0^2 = 10^{-2} {\cal T}_1^2$ and
$\Pi^2 = 10^{-6}\beta_0^2$.
From the bottom, the lines correspond to
$t=0,50,100,...,300,310,320,...,370,374$.
The overall evolution is very similar to the pure tachyon case,
but is lagged a little bit due to the nonvanishing electric flux.
This case confronts a caustic in the end.
}
\label{FIG2}
\end{figure}

Since we are interested in the tachyon evolution
in comparison with the pure tachyon case,
first we fix the value of $\beta_0$ and vary $\Pi$
in the range (\ref{Pirange}).
When $\Pi$ is very small, we observe from numerical calculations
that the tachyon evolves in a very similar way to the
pure tachyon case.
In Fig.~\ref{FIG2}, we show the numerical results for
$\beta_0 = 0.1{\cal T}_1$ and $\Pi = 10^{-3}\beta_0$ (i.e., $E_0 = 10^{-3}$).
The tachyon field grows in time in the bulk,
and around the kink/anti-kink at late times
it exhibits a similar caustic behavior
to what happened in the $\Pi =0$ case.
When the electric flux is small,
the fundamental string cannot
properly modulate the tachyon flow to the kink direction.
This picture is common for all the values of $\beta_0$,
and even for $\beta_0^2>1$ which was a prohibited range
for the pure tachyon case.

\begin{figure}[!h]
\centerline{\epsfig{figure=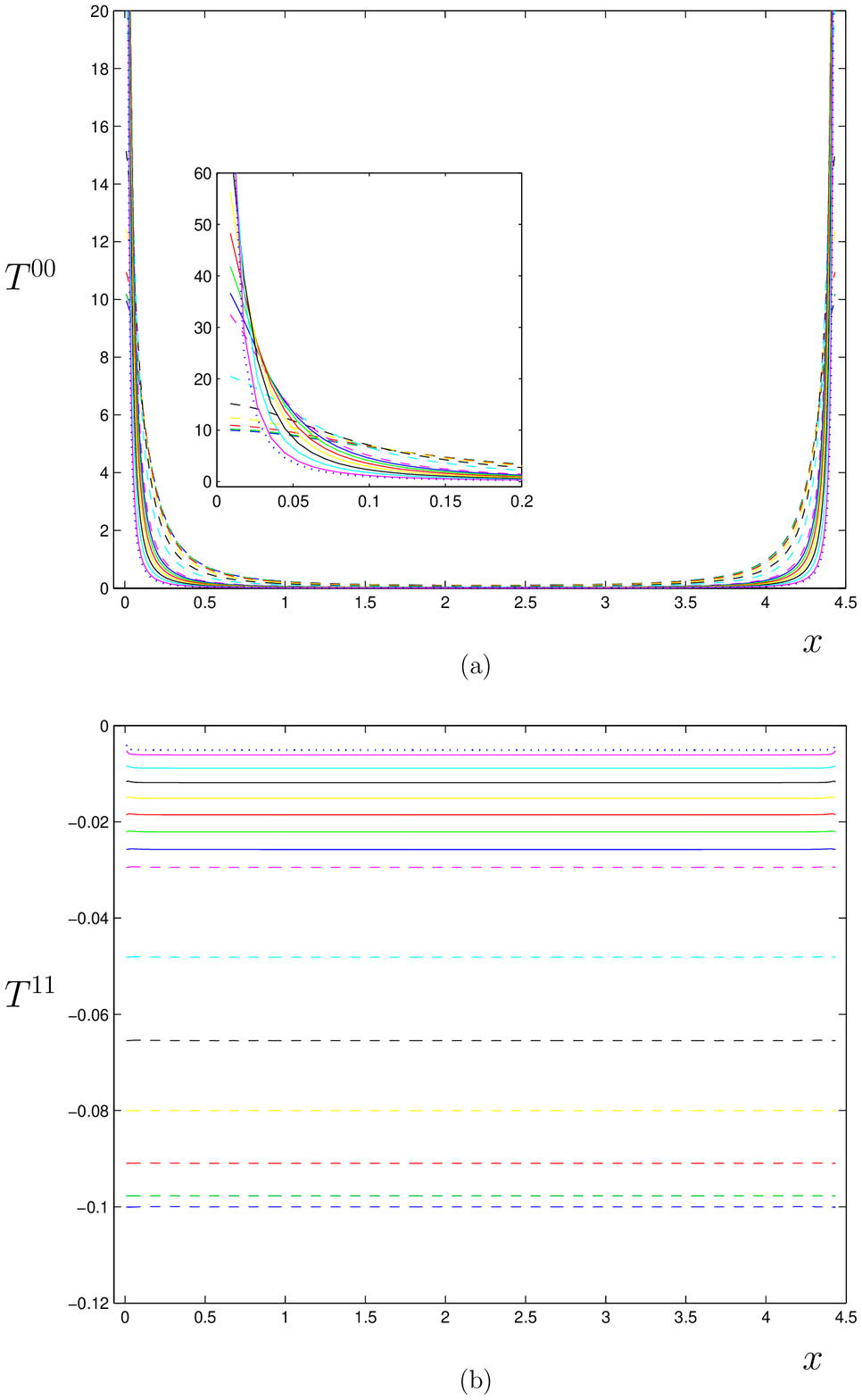,height=180mm}}
\caption{
Plot of (a) energy density $T^{00}$
and (b) pressure $T^{11}$.
As the tachyon evolves,
the energy flows from the bulk to the kink
and the kink becomes thin.
(See the zoom-in plot of the kink region.)
The pressure gradually increases to zero from the below.
}
\label{FIG3}
\end{figure}

As time elapses, the energy flows from the bulk region
to the kink/anti-kink (See Fig.~\ref{FIG3}).
The energy density (\ref{T00}) in Hamiltonian representation
is given by
\bear\label{HH}
\rho = {\cal H} = \sqrt{(1+{T'}^2)(\Pi^2 +\Pi_T^2 +V^2)},
\eear
where the conjugate momentum of the tachyon field $T$ is
\bear
\label{PiT}
\Pi_T = {\partial {\cal L} \over \partial \dot{T}} = \beta\dot{T}.
\eear
The plot of $\rho = T^{00}$ shows that the energy is lumped
at the kink/anti-kink,
and the corresponding lower dimensional D0/$\bar{\rm D}0$-branes
become thin due to the nonvanishing momentum flow toward the kink location.
As time elapses further, the evolution steps into the initial
stage of the caustic formation in a very similar manner
discussed in the pure tachyon case,
beyond which numerical calculations cannot be performed.

\begin{figure}[!h]
\centerline{\epsfig{figure=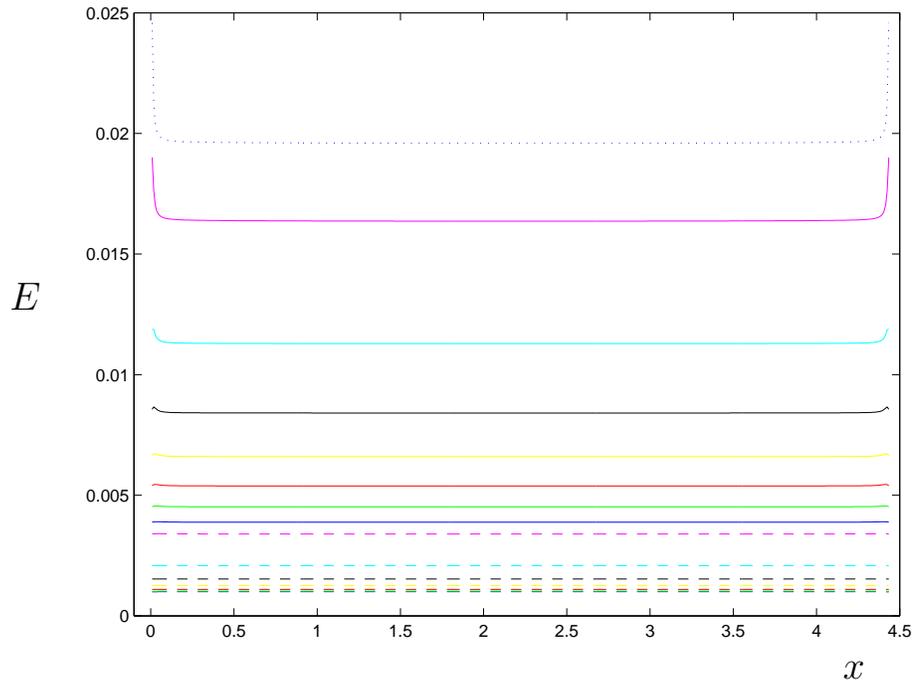,height=90mm}}
\caption{
Plot of electric field $E$.
$E$ grows more or less homogeneously in the bulk region,
and peaks near the kink/anti-kink.
}
\label{FIG4}
\end{figure}

The pressure increases pretty homogeneously to zero from the below
as shown in Fig.~\ref{FIG3}.
The electric field which initially began from a constant value
increases more or less homogeneously in the bulk region,
but it peaks up at the kink/anti-kink region at late times
as shown in Fig.~\ref{FIG4}.

Similarly to the case of the pure tachyon in Section 3.1, we consider
the decay products analytically. Substituting the static solution
(\ref{staticT}) into the energy density relation (\ref{T00}),
we obtain
\bear\label{T00-1}
T^{00} = \Pi E_0 + \frac{E_0}{\Pi}\,\frac{{\cal T}_1^2}{
1+ \left(\frac{E_0^2 {\cal T}_0^2}{\Pi^2(1-E_0^2)}-1\right)\,
\sin^2\left(\frac{x}{\xi}\right)},
\eear
where $\xi= \sqrt{2}/\sqrt{1-E_0^2}$.
Integrating over a half period centered at the kink, we obtain the initial
total energy generated by the kink~\cite{Kim:2003in},
\bear\label{Ekink}
E_{{\rm kink}} =
\int_{-\frac{\xi\pi}{2}}^{\frac{\xi\pi}{2}} dx\, T^{00}
=  \frac{\sqrt{2}\pi\Pi^2}{\sqrt{\beta_0^2-\Pi^2}} + {\cal T}_0,
\eear
where $\beta_0=\Pi/E_0$ and ${\cal T}_0 = \sqrt{2}\pi {\cal T}_1$.
The terms on the right hand side of Eq.~(\ref{Ekink}) denote
the string charge over a half circle and the tension of the D0-brane respectively.
The total energy of the kink profile depends
on both $\Pi$ and $\beta_0$.
The thickness of the kink can also be deduced from
the slope of the static solution,
\bear\label{slop}
T_0(x) \sim \frac{1}{\beta_0}\,\sqrt{{\cal T}_1^2 - (\beta_0^2-\Pi^2)}\, x.
\eear
For a given $\Pi$, once the tachyon is unstable
and experiences changes in its thickness,
the evolution would not be in such a way
as to follow other static configurations
which are described by different $\beta_0$'s;
the total energy $E_{{\rm kink}}$ is not conserved.
Being different from the pure tachyon case in this sense,
it is hard to imagine what the final decay products would be.
In particular, for our current system in which the electric field $E$
evolves inhomogeneously near the kink,
and which seems to accompany a caustic behavior at late times,
it is impossible to discuss analytically the decay products at the final stage.

\subsubsection{Stable Solution}

As the value of $\Pi$ is increased,
the electric flux on the fundamental string
suppresses more efficiently the tachyon flow toward the kink as well as
the tachyon decay in the bulk.
Above some critical value of $\Pi$, therefore,
we expect that the flux completely blocks
the tachyon evolution and makes the tachyon stable.
The existence of the electric flux (the ${T'}^2\Pi^2$ term)
takes parts in the energy density~(\ref{HH}),
which presumably suppress the kinetic portion of the tachyon field
for a given system.

\begin{figure}[!h]
\centerline{\epsfig{figure=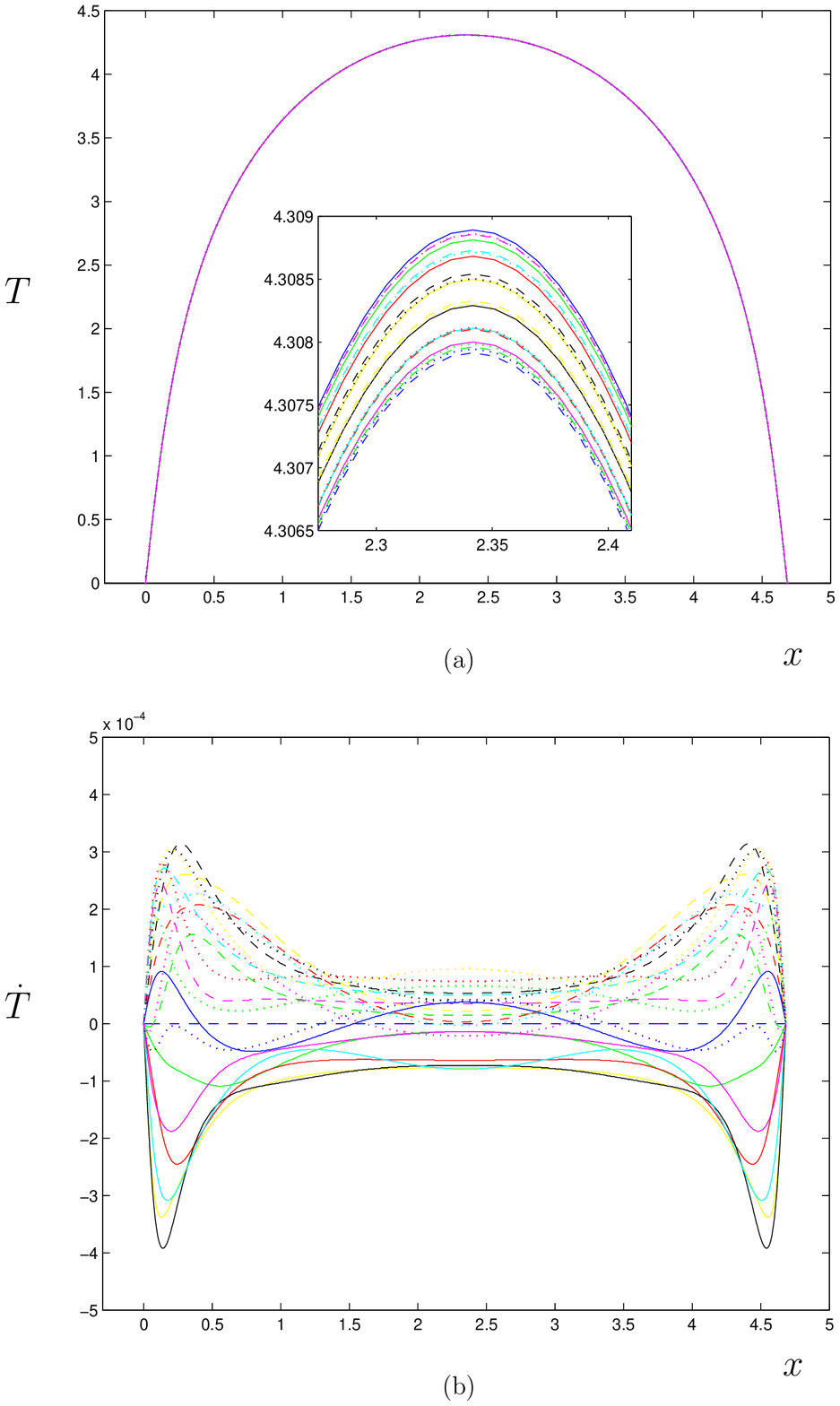,height=180mm}}
\caption{
Plot of (a) $T(t,x)$ and (b) $\dot{T}(t,x)$ for the tachyon-plus-electric
field case with $\beta_0^2 = 10^{-2} {\cal T}_1^2$ and $\Pi^2
= 10^{-1}\beta_0^2$.
The lines correspond to
$t=0,30,60,...,600$, but are distinguishable only in the zoom-in plot of
the top region.
The tachyon field is very stable and exhibits only very small oscillations.
Although we have not plotted enough time slices,
the top part of the tachyon configuration
experiences about 20 oscillations till $t=600$.
In (b), $\dot{T}$ also oscillates about zero stably.
(Depending on the values of $\beta_0$ and $\Pi$,
$\dot{T}$ can be less wiggly.)
}
\label{FIG5}
\end{figure}

\begin{figure}[!h]
\centerline{\epsfig{figure=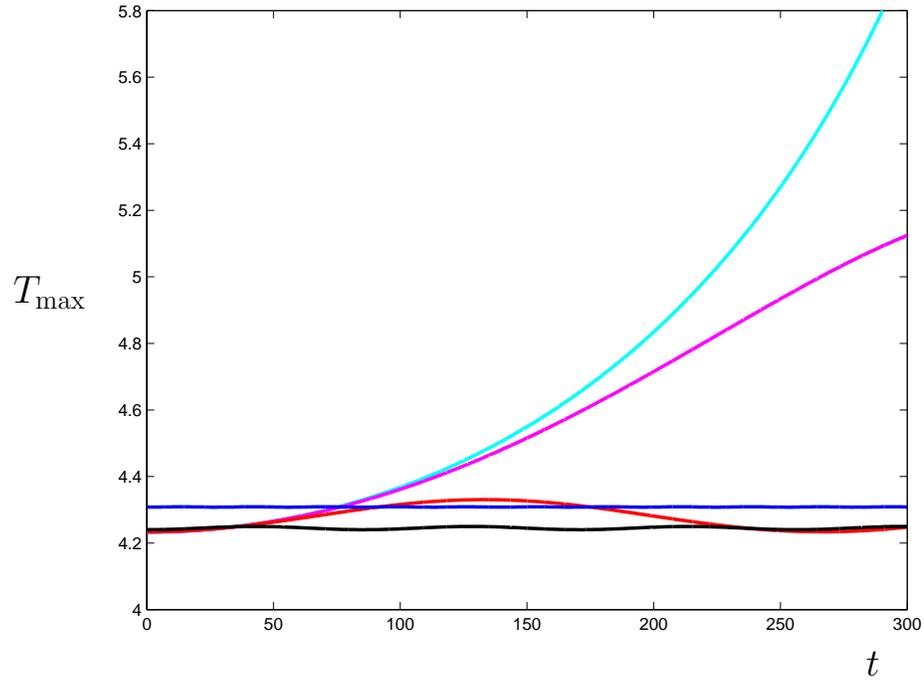,height=90mm}}
\caption{
Plot of $T_{\rm max}$ for $\beta_0^2 = 10^{-2} {\cal T}_1^2$
and $\Pi^2 = 10^{-6}, 10^{-4}, 10^{-3}, 10^{-2}, 10^{-1} \beta_0^2$.
For subcritical values of $\Pi$, $T_{\rm max}$ grows monotonically.
For supercritical values of $\Pi$, $T_{\rm max}$ becomes stable.
}
\label{FIG5.2}
\end{figure}

From numerical calculations, we observe that  the critical value
ranges around $\Pi_c(\beta_0) \sim 0.1\beta_0$, or less.
The numerical coefficient also depends on $\beta_0$.
In Fig.~\ref{FIG5}, we show the numerical results for
$\beta_0 = 0.1 {\cal T}_1$ and $\Pi = \sqrt{0.1}\beta_0$.
The tachyon remains very close to the initial static solution
with only small oscillations (these are visible in the zoom-in plot only).
The velocity $\dot{T}$ oscillates about zero, which also indicates
the stability of tachyon.
In Fig.~\ref{FIG5.2}, we plotted the evolution of the maximum
value of $T$ ($T_{\rm max}$) for several values of $\Pi$.

\begin{figure}[!h]
\centerline{\epsfig{figure=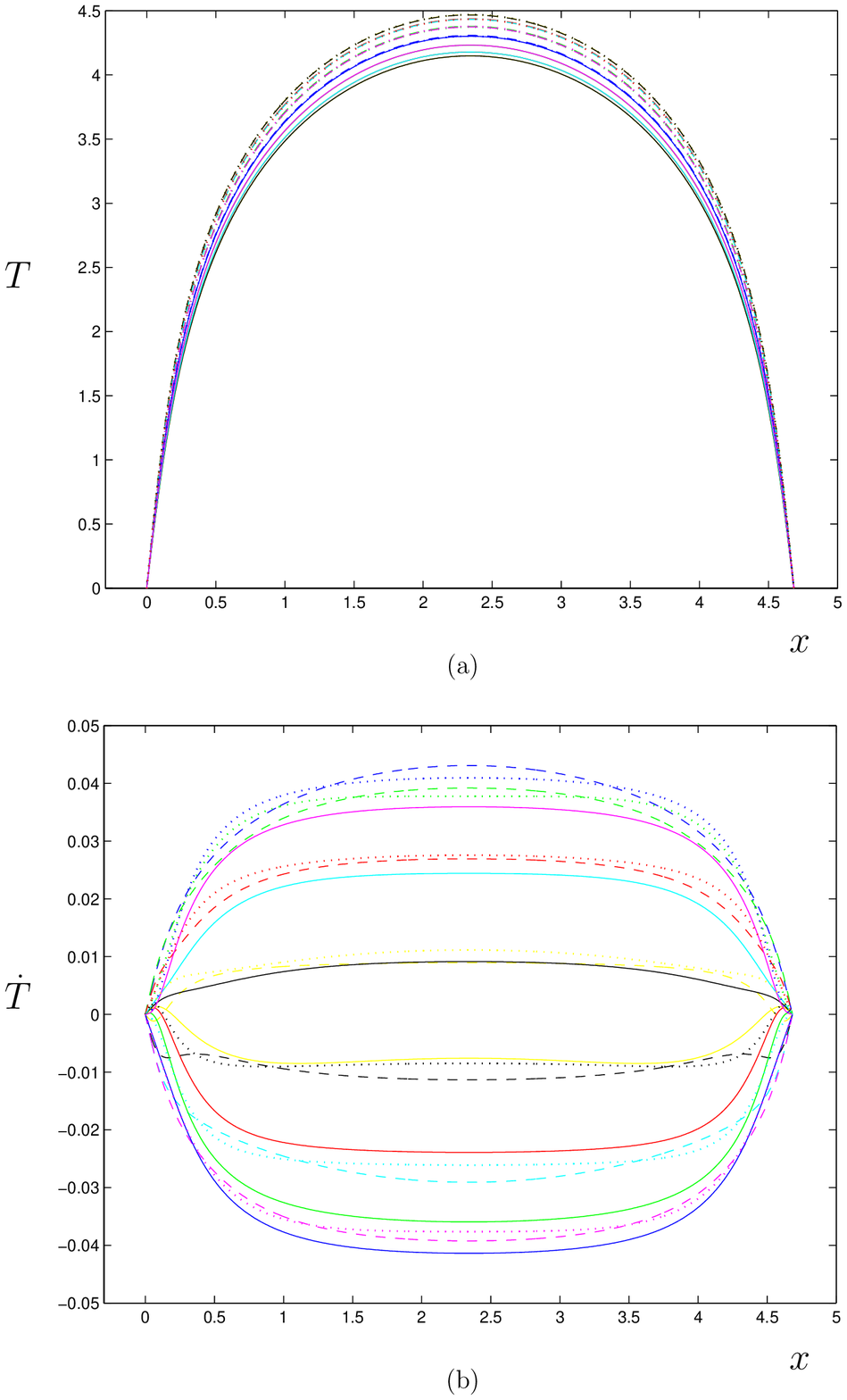,height=180mm}}
\caption{
Plot of (a) $T(t,x)$ and (b) $\dot{T}(t,x)$
for the case of a larger perturbation
with an initial velocity $\dot{T}(0,x) = 0.01T(0,x)$.
The other parameters are the same as in Fig.~\ref{FIG5}.
The tachyon field oscillates stably
about the static solution as in the minimal perturbation case,
but with a larger amplitude.
}
\label{FIG6}
\end{figure}

For numerical calculations so far,
we have adopted the initial condition (\ref{IC2})
which implies only a minimal perturbation.
Therefore, one may question
what would happen if larger perturbations are applied
to the stable solutions.
The conclusion is that the stable solution remains
still stable unless the perturbations are unreasonably large.
We have performed numerical calculations with an initial condition,
$\dot{T}(t=0,x) = c_1 \times T(t=0,x)$,
by varying the constant $c_1$ in a reasonable range ($|c_1| \ll 1$).
In Fig.~\ref{FIG6}, we plotted the tachyon evolution for $c_1 = 0.01$
with the same values of the other parameters.
The tachyon field remains still stable,
although the oscillation amplitude gets larger
than that of the minimal perturbation case.
For larger $c_1$'s, the oscillation amplitudes are larger.

According to our numerical results, the critical electric flux
depends on the value of $\beta_0$, $\Pi_c(\beta_0) \sim 0.1 \beta_0$.
For a given $\beta_0$, the stability of the kink is
guaranteed if $\Pi \geq \Pi_c$.
Conversely, for a given $\Pi$,
$\beta_0$ should range as
\bear\label{ranbeta}
\Pi < \beta_0 \lesssim 10 \Pi_c .
\eear
The initial static kink with $\beta_0$ beyond the
above range, is subject to instability.
The slope of the kink,
$\sqrt{{\cal T}_1^2-(\beta_0^2-\Pi^2)}/\beta_0$,
also has a constrained range for the stability accordingly;
the thickness of the lower dimensional D-brane does so,
because it is evaluated roughly as the inverse of the slope.

\section{Semi-analytic Approach}

In the previous section,
we observed from {\it numerical} calculations
that the static tachyon configuration is
stable when the electric flux is sufficiently large.
For small or zero electric flux, however,
the static configuration is no longer stable; the tachyon
rolls down the potential and the caustic forms at the kink.
In this section, we would like to capture the stabilization story
more conceptually mainly by analyzing the field equation.
The analysis will be a bit technical.

The tachyon field equation (\ref{te2}) can be arranged as
\bear\label{te7}
(1+{T'}^2) \ddot T = (1-\dot T^2)\left[T'' +(1+{T'}^2)\,B(T)\right]
+ 2\dot T T'\dot T' + \dot T^2{T'}^2B(T) ,
\eear
where
\bear\label{BT}
B(T) \equiv
\left\{%
\begin{array}{ll}
{1\over \sqrt{2}}\tanh\left(\frac{T}{\sqrt{2}}\right), & \hbox{for $\Pi = 0$,}\\
\frac{{\cal T}_1^2
\mbox{sech}^2\left(\frac{T}{\sqrt{2}}\right)
\tanh\left(\frac{T}{\sqrt{2}}\right)}{\sqrt{2}
\left[\Pi^2 + {\cal T}_1^2\mbox{sech}^2\left(\frac{T}{\sqrt{2}}\right)
\right]}, & \hbox{for $\Pi \neq 0$.} \\
\end{array}%
\right.
\eear
It is enough to consider only one patch between
the kink ($x = x_0 = 0$) and
the anti-kink ($x=x_1 = \sqrt{2}\pi\beta_0 / \sqrt{\beta_0^2-\Pi^2}$)
as we did in numerical calculations.
Let us call the location of the maximum of $T$ as $x_{1/2}$ ($=x_1/2$).
Equalizing the square-bracket part in Eq.~(\ref{te7}) to zero
corresponds to the static equation.
We adopted its solution as the initial configuration.
Applying the minimal perturbation discussed in the previous section,
we are now interested in how $\ddot T$ behaves in order to check the stability.
If its signature remains unchanged,
$T$ will grow/decrease only;
the tachyon is unstable.
In order to have a stably oscillating tachyon configuration,
the signature of $\ddot T$ should keep changing during the evolution.
It is sufficient to look up this behavior only in some parts of the patch,
so we shall do in the bulk region
where the evolution is gentle.

\subsection{Pure Tachyon Case $(\Pi =0)$}
Suppose that the tachyon field begins to evolve initially
with a tiny positive velocity $\dot T(t=0,x) >0$
from the static configuration.
The Noether momentum along the $x$-direction,
\bear\label{mom1}
P_1= - \beta \dot T T',
\eear
is always negative (positive) in the region $0<x<x_{1/2}$ ($x_{1/2}<x<x_1$).
Therefore, the energy density flows to the kink (anti-kink) accordingly,
and there grows the slope of $T$ as well.
And the corresponding term, $2\dot T T'\dot T' + \dot T^2{T'}^2B(T)$,
in Eq.~(\ref{te7}) is always positive.

The remaining factor in Eq.~(\ref{te7})
in determining the change in $\ddot T$
is now the square-bracket part
\bear\label{inkink1}
\left[T'' + \frac{1+{T'}^2}{\sqrt{2}} \tanh
\left(\frac{T}{\sqrt{2}}\right)\right].
\eear
Initially this is zero since we adopted the static solution
as the initial condition.
As time elapses, we observed from numerical calculations
that the curvature $T''$ remained more or less unchanged
in negative in the bulk while $T$ grows.
Therefore, the above term~(\ref{inkink1}) remains positive
since its second term is a growing function of $T$.

As a whole, all the terms on the right hand side of Eq.~(\ref{te7})
remain positive, which keeps $\ddot T$ positive;
the tachyon is unstable to grow in the bulk region and to form caustic
about the kink and the anti-kink.

The tachyon velocity $\dot T$ does not exceed one
in the middle of the bulk region, which can be
understood from the following.
The slope $T'$ and the curvature $T''$ in this region
is relatively small,
so we can approximate the tachyon field equation to
a homogeneous one
\bear\label{homeq}
\ddot T = \frac1{\sqrt{2}}(1-\dot
T^2)\tanh\left(\frac{T}{\sqrt{2}}\right).
\eear
The solutions to this equation are given by~\cite{Kim:2005pz},
\bear\label{homsol}
T(t) = \left\{\begin{array}{c} \sqrt{2}\sinh^{-1}\left[A_{{\rm c}}
\cosh\left(\frac{t}{\sqrt{2}}\right)\right]  \\
\sqrt{2}\sinh^{-1}\left[A_{{\rm e}}
\exp \left(\frac{t}{\sqrt{2}}\right)\right]  \\
\sqrt{2}\sinh^{-1}\left[A_{{\rm s}}
\sinh\left(\frac{t}{\sqrt{2}}\right)\right]
\end{array}\right.,
\eear
where the constants $A_{{\rm c}}$ and $A_{{\rm s}}$
depend on the initial energy density and are less than $\sqrt{2}$
while $A_{{\rm e}}$ is an arbitrary constant.
Then $|\dot T|$ remains less than one for all of the solutions
in Eq.~(\ref{homsol}) and approaches one at infinity.

\subsection{Tachyon with Electric Flux ($\Pi \neq 0$)}
\begin{figure}[!h]
\centerline{\epsfig{figure=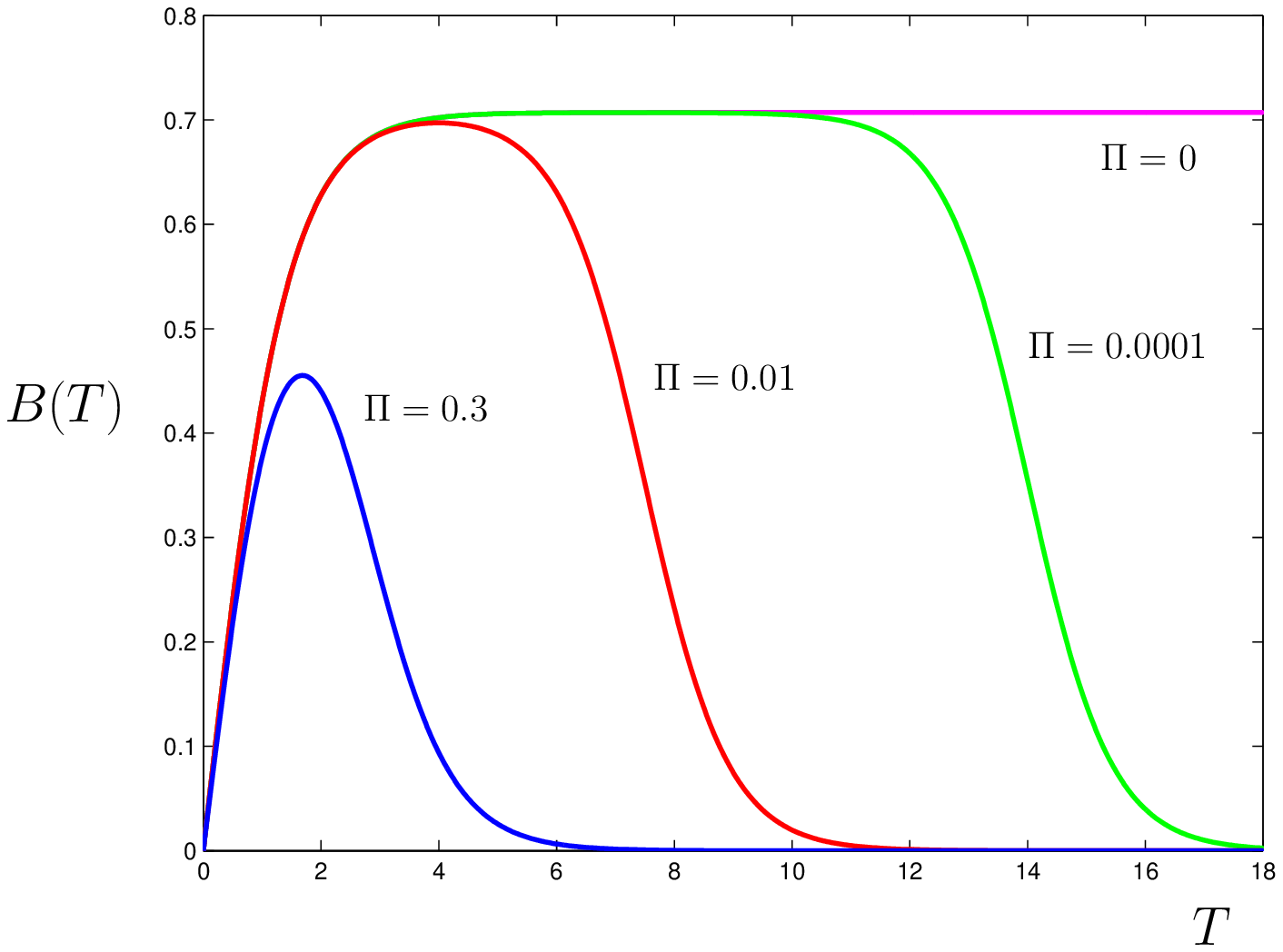,height=90mm}}
\caption{Profiles of $B(T)$ for various $\Pi$.}
\label{FIG7}
\end{figure}

If the electric flux is turned on,
the only difference in the field equation~(\ref{te7})
is $B(T)$ defined in Eq.~(\ref{BT}).
Unlike the pure tachyon case,
$B(T)$ is now a bump-like function of $T$ of which maximum is
at
\bear\label{Tmax}
T= T_{{\rm max}} \equiv \sqrt{2} \mbox{sech}^{-1}\left[\frac{\sqrt{(9\Pi^4
+ 8\Pi^2{\cal T}_1^2)^{1/2}- 3 \Pi^2}}{\sqrt{2 {\cal T}_1^2}}\right].
\eear
We plotted $B(T)$ for several values of $\Pi$ in Fig.~\ref{FIG7}.

The square-bracket part
\bear\label{inkink2}
\left[T'' + (1+{T'}^2)B(T)\right]
\eear
which is initially set to zero
can become negative
if $T$ is on the descending region of $B(T)$.
Then, it can possibly happen that
the acceleration $\ddot T$ becomes negative.
If the decelerating state ($\ddot T <0$) maintains,
the velocity will change its signature
and become negative, $\dot T <0$.
Then $T$ decreases and rolls back up  the right hill of $B(T)$.
The square-bracket (\ref{inkink2}), $\ddot T$, and $\dot T$
evolve in an opposite way to that during descending.
$T$ will eventually change its direction in the end.
This process keeps $T$ oscillating about a certain configuration,
and thus the tachyon becomes stable.
The stabilization in this way is more probable for larger $\Pi$
since the maximum of $B(T)$ is lower;
a certain amount of electric flux is required
in order to halt the tachyon rolling down the potential
and the flow to the kink.
This agrees with our numerical results obtained in the previous section.

\subsection{Caustic Formation Revisited}
Before we close this section,
let us consider the tachyon flow and the caustic formation in the kink region.
The quantity $-X$ in Eq.~(\ref{XX}) should be positive
for a physical system.
However, when a caustic formation is accompanied ($\Pi<\Pi_c$),
it was observed from numerical calculations that
the value of $-X$  at the adjacent position to the kink
falls to zero and then becomes negative.
This behavior of $-X$ occurs just before the numerical calculation
diverges due to the large slope of $T$.
This is very similar to the pure tachyon result in Ref.~\cite{Cline:2003vc}
where the authors used a different tachyon potential.

Following the techniques in Refs.~\cite{Felder:2002sv,Cline:2003vc},
we can estimate the critical time at which the fluid element
at the adjacent position reaches the kink.
Once $-X$ becomes zero at a position,
assume that the tachyon evolution is constrained by
\bear\label{X2}
-X &=& 1-\dot T^2 + T^{'2}-E^2=\frac{{\cal T}_1^2 V^2}{\Pi^2 +{\cal T}_1^2 V^2}
\left(1- \dot T^2 + {T'}^2\right)=0,
\eear
where we used Eq.~(\ref{E1}) in the second step.
As ${\cal T}_1^2 V^2/(\Pi^2 + {\cal T}_1^2 V^2)$ is nonzero,
for $\Pi <\Pi_c$,
the system is governed by the equation
\bear\label{PP}
 1- \dot T^2 + {T'}^2 =0,
\eear
which is exactly the same one for the pure tachyon case.
Then the results in Refs.~\cite{Felder:2002sv,Cline:2003vc}
are applied exactly in the same way as follows.

This first order partial differential equation was solved analytically
by the method of characteristics~\cite{Felder:2002sv}.
We introduce a parameter $q$ which satisfies $x(q,t_i)=q$, $t(q,t_i)=0$
, and $T(q,t_i)=T_i(q)$ at the initial moment $t_i$ (we set $t_i=0$).
Then the solutions to the equation~(\ref{PP}) are
\bear
x_c(q,t)&=& q-\frac{T_{i,q}}{\sqrt{1+T_{i,q}^2}}\,t ,\\
T(q,t)&=& T_i(q)+\frac{t}{\sqrt{1+T_{i,q}^2}},
\eear
where $T_{i,q}\equiv {\partial T_i}/{\partial q}$ and $x_c(q,t)$ is the
characteristic curve of $q$ at $t$.
After a finite time $t= q\sqrt{1+T_{i,q}^2}/T_{i,q}$, $x_c$ will cross
$x=0$ with a nonzero value of tachyon field,
$T(q,t) = T_i(q) + q/T_{i,q}$.
Finally the tachyon field becomes multi-valued there, and
its slope diverges~\cite{Cline:2003vc}.
The caustic forms firstly
at the minimum of $t= q\sqrt{1+T_{i,q}^2}/T_{i,q}$.

\section{Conclusions}

We investigated the numerical evolution of an inhomogeneous D1-brane
with a string charge density.
We considered a DBI-type action
which contains a tachyon field $T$ with a run-away potential
$V=1/\cosh (T/\sqrt{2})$ and a gauge field $A_1$
in a flat (1+1) dimensional spacetime.
The target space that we considered is a compactified circle $S^1$
with a fixed radius.
The tachyon field presents an inhomogeneous kink and anti-kink array
on the circle.
The fundamental string which we considered
induces an electric field $E$ on the D1-brane.
The electric flux $\Pi$ (the string charge density)
is solved to be constant while the electric field
is let to vary both in time and in space.

There exists a static solution when the electric field is constant,
$E=E_0$~\cite{Lambert:2003zr,Kim:2003in}.
We focus our interest on the case of $ 0 \leq E_0 < 1$
for which the tachyon profiles represent a kink-anti-kink array.
For the pure tachyon case ($E_0 =0$),
it had been studied in Refs.~\cite{Felder:2002sv,Cline:2003vc}
that the tachyon field is unstable
to form a caustic at the location of the kink, or the anti-kink;
the tachyon field becomes double-valued at a finite time,
and thus the lower dimensional D-brane becomes thin.

For the initial configuration in evolving the tachyon field,
we adopt the static solution of which configuration is determined
by two parameters, the electric flux $\Pi$ and the electric field $E_0$
(or $\beta_0=\Pi/E_0$).
The compactification radius is then fixed to $\sqrt{2}/\sqrt{1-E_0^2}$.
The electric flux $\Pi$ remains constant during the evolution as a solution
to the gauge field equation, while $E$ ($\beta$) is allowed change.
Applying a minimal perturbation on the initial configuration,
we evolved the tachyon field.

When the string charge density $\Pi$ is zero or sufficiently small,
the tachyon is unstable regardless of the initial pressure
of the tachyon kink, $-\beta_0$.
$T$ rolls down the potential $V$, and its slope gets large
at the kink and the anti-kink.
Although it is never possible to reach numerically the infinite-slope state,
we expect that the tachyon configuration approaches the situation
of caustic formation in a similar way discussed in Ref.~\cite{Cline:2003vc}.
According to the Sen's proposal in \cite{Sen:2002vv},
the final state can be interpreted as a thin D0-$\bar{\rm D}0$ pair
located diametrically opposite points on the
circle~\cite{Sen:1998tt,Sen:1998ex,Sen:2003zf}.

When $\Pi$ is sufficiently large,
the tachyon becomes stabilized.
The initial static solution remains stable even when larger
perturbations are applied.
The fundamental string connecting D0- and $\bar{\rm D}0$-branes
modulates the tachyon flow and provides the stability.
The stabilization is achieved
when the electric flux is larger than some critical
value $\Pi > \Pi_c$ where $\Pi_c$ is a function of the initial $\beta_0$,
and numerically it is $\Pi_c \sim 0.1\beta_0$. The stable kink
with $\Pi > \Pi_c$ is identified as the lower dimensional
stable thick D-brane with thickness
$\sim \beta_0/\sqrt{{\cal T}_1 - \beta_0^2}$.

It is not sufficient, however, to fully discuss the stability of
the resulting thick D0-$\bar {\rm D}0$ pair with our limited boundary
and initial conditions. To clarify the properties of the system,
one needs more investigations with various boundary and initial
conditions. For example, if one imposes a deformed initial
velocity of the tachyon and an appropriate boundary condition, one
would be able to observe pair annihilation of kink and anti-kink
which is related to the instability due to
D0 and $\bar{\rm D}0$ charges.

Another example will be to consider
varying radius of $S^1$ in time,
or to consider the evolution on $R_1$.
In this case, the period of the tachyon kink
which was initially set to $2\sqrt{2}\pi/\sqrt{1-E_0^2}$,
may change in time.
One possible way of evolution is
to track the successive static solutions  of different periods.
The electric field then evolves homogeneously.

Our numerical study in this work was the first step to
investigating the instability of the tachyon-plus-electric flux system,
with the conditions which preserve involved symmetries maximally.
In the future, we hope our work is extended to other various directions.

\section*{Acknowledgements}
We are grateful to Gary Felder, Gungwon Kang, Chanju Kim,
Yoonbai Kim, and Piljin Yi for very useful discussions.
This work was supported by the BK 21 project of the
Ministry of Education and Human Resources Development, Korea
(I.C. and C.O.L.),
by the Astrophysical Research Center
for the Structure and Evolution of the Cosmos (ARCSEC)
and the grant No.~R01-2006-000-10965-0 from the Basic Research Program
of the Korea Science \& Engineering Foundation (I.C.),
and by the Science Research Center Program of
the Korea Science and Engineering Foundation through
the Center for Quantum Spacetime (CQUeST) of
Sogang University with grant number R11-2005-021 (O.K.).

\clearpage
\clearpage
\clearpage
\clearpage
\clearpage
\clearpage
\clearpage


\begin{thebibliography}{99}

\bibitem{Sen:2002vv}
  A.~Sen,
  ``Time evolution in open string theory,''
  JHEP {\bf 0210}, 003 (2002)
  [arXiv:hep-th/0207105].


\bibitem{Larsen:2002wc}
  F.~Larsen, A.~Naqvi and S.~Terashima,
  ``Rolling tachyons and decaying branes,''
  JHEP {\bf 0302}, 039 (2003)
  [arXiv:hep-th/0212248].


\bibitem{Rey:2003zj}
  S.~J.~Rey and S.~Sugimoto,
  ``Rolling of modulated tachyon with gauge flux and emergent fundamental
  string,''
  Phys.\ Rev.\ D {\bf 68}, 026003 (2003)
  [arXiv:hep-th/0303133].


\bibitem{Ishida:2003cj}
  A.~Ishida and S.~Uehara,
  ``Rolling down to D-brane and tachyon matter,''
  JHEP {\bf 0302}, 050 (2003)
  [arXiv:hep-th/0301179].


\bibitem{Felder:2002sv}
  G.~N.~Felder, L.~Kofman and A.~Starobinsky,
  ``Caustics in tachyon matter and other Born-Infeld scalars,''
  JHEP {\bf 0209}, 026 (2002)
  [arXiv:hep-th/0208019].


\bibitem{Mukohyama:2002vq}
  S.~Mukohyama,
  ``Inhomogeneous tachyon decay, light-cone structure and D-brane network
  problem in tachyon cosmology,''
  Phys.\ Rev.\ D {\bf 66}, 123512 (2002)
  [arXiv:hep-th/0208094].


\bibitem{Berkooz:2002je}
  M.~Berkooz, B.~Craps, D.~Kutasov and G.~Rajesh,
  ``Comments on cosmological singularities in string theory,''
  JHEP {\bf 0303}, 031 (2003)
  [arXiv:hep-th/0212215].


\bibitem{Cline:2003vc}
  J.~M.~Cline and H.~Firouzjahi,
  ``Real-time D-brane condensation,''
  Phys.\ Lett.\ B {\bf 564}, 255 (2003)
  [arXiv:hep-th/0301101].


\bibitem{Felder:2004xu}
  G.~N.~Felder and L.~Kofman,
  ``Inhomogeneous fragmentation of the rolling tachyon,''
  Phys.\ Rev.\ D {\bf 70}, 046004 (2004)
  [arXiv:hep-th/0403073].


\bibitem{Barnaby:2004dp}
  N.~Barnaby and J.~M.~Cline,
  ``Creating the universe from brane-antibrane annihilation,''
  Phys.\ Rev.\ D {\bf 70}, 023506 (2004)
  [arXiv:hep-th/0403223].

\bibitem{Barnaby:2004nk}
  N.~Barnaby,
  ``Caustic formation in tachyon effective field theories,''
  JHEP {\bf 0407}, 025 (2004)
  [arXiv:hep-th/0406120].


\bibitem{Panigrahi:2004qr}
  K.~L.~Panigrahi,
  ``D-brane dynamics in Dp-brane background,''
  Phys.\ Lett.\ B {\bf 601}, 64 (2004)
  [arXiv:hep-th/0407134].


\bibitem{Saremi:2004yd}
  O.~Saremi, L.~Kofman and A.~W.~Peet,
  ``Folding branes,''
  Phys.\ Rev.\ D {\bf 71}, 126004 (2005)
  [arXiv:hep-th/0409092].


\bibitem{Canfora:2005mt}
  F.~Canfora,
  ``A note on tachyon dynamics,''
  Phys.\ Lett.\ B {\bf 625}, 277 (2005)
  [arXiv:gr-qc/0508082].


\bibitem{Kwon:2006wh}
  O-K.~Kwon and C.~O.~Lee,
  ``Late time behaviors of an inhomogeneous rolling tachyon,''
  Phys.\ Rev.\ D {\bf 73}, 126001 (2006)
  [arXiv:hep-th/0601236].


\bibitem{Sen:2002nu}
  A.~Sen,
  ``Rolling tachyon,''
  JHEP {\bf 0204}, 048 (2002)
  [arXiv:hep-th/0203211].


\bibitem{Sen:2002in}
  A.~Sen,
  ``Tachyon matter,''
  JHEP {\bf 0207}, 065 (2002)
  [arXiv:hep-th/0203265].


\bibitem{Sen:2002an}
  A.~Sen,
  ``Field theory of tachyon matter,''
  Mod.\ Phys.\ Lett.\ A {\bf 17}, 1797 (2002)
  [arXiv:hep-th/0204143].


\bibitem{Sugimoto:2002fp}
  S.~Sugimoto and S.~Terashima,
  ``Tachyon matter in boundary string field theory,''
  JHEP {\bf 0207}, 025 (2002)
  [arXiv:hep-th/0205085]~;
  J.~A.~Minahan,
  ``Rolling the tachyon in super BSFT,''
  JHEP {\bf 0207}, 030 (2002)
  [arXiv:hep-th/0205098]~;
  A.~Ishida, Y.~Kim and S.~Kouwn,
  ``Homogeneous rolling tachyons in boundary string field theory,''
  Phys.\ Lett.\ B {\bf 638}, 265 (2006)
  [arXiv:hep-th/0601208].


\bibitem{Gibbons:2002tv}
  G.~Gibbons, K.~Hashimoto and P.~Yi,
   ``Tachyon condensates, Carrollian contraction of Lorentz group, and
  JHEP {\bf 0209}, 061 (2002)
  [arXiv:hep-th/0209034].


  \bibitem{Kim:2003he}
C.~Kim, H.~B.~Kim, Y.~Kim and O-K.~Kwon,
``Electromagnetic string fluid in rolling tachyon,''
JHEP {\bf 0303}, 008 (2003)
[arXiv:hep-th/0301076].


\bibitem{Sen:2003tm}
  A.~Sen,
  ``Dirac-Born-Infeld action on the tachyon kink and vortex,''
  Phys.\ Rev.\ D {\bf 68}, 066008 (2003)
  [arXiv:hep-th/0303057].


\bibitem{Lambert:2003zr}
  N.~Lambert, H.~Liu and J.~Maldacena,
  ``Closed strings from decaying D-branes,''
  arXiv:hep-th/0303139.


\bibitem{Brax:2003rs}
  P.~Brax, J.~Mourad and D.~A.~Steer,
  ``Tachyon kinks on non BPS D-branes,''
  Phys.\ Lett.\ B {\bf 575}, 115 (2003)
  [arXiv:hep-th/0304197]~;
  E.~J.~Copeland, P.~M.~Saffin and D.~A.~Steer,
  ``Singular tachyon kinks from regular profiles,''
  Phys.\ Rev.\ D {\bf 68}, 065013 (2003)
  [arXiv:hep-th/0306294];
  D.~Bazeia, R.~Menezes and J.~G.~Ramos,
  ``Regular and periodic tachyon kinks,''
  Mod.\ Phys.\ Lett.\ A {\bf 20}, 467 (2005)
  [arXiv:hep-th/0401195].


\bibitem{Kim:2003in}
  C.~Kim, Y.~Kim and C.~O.~Lee,
  ``Tachyon kinks,''
  JHEP {\bf 0305}, 020 (2003)
  [arXiv:hep-th/0304180];
  C.~Kim, Y.~Kim, O-K.~Kwon and C.~O.~Lee,
  ``Tachyon kinks on unstable Dp-branes,''
  JHEP {\bf 0311}, 034 (2003)
  [arXiv:hep-th/0305092].


\bibitem{Afonso:2006ws}
  V.~Afonso, D.~Bazeia and F.~A.~Brito,
  ``Deforming tachyon kinks and tachyon potentials,''
  JHEP {\bf 0608}, 073 (2006)
  [arXiv:hep-th/0603230].


\bibitem{Kim:2006mg}
  C.~Kim, Y.~Kim, O-K.~Kwon and H.~U.~Yee,
  ``Tachyon kinks in boundary string field theory,''
  JHEP {\bf 0603}, 086 (2006)
  [arXiv:hep-th/0601206].


\bibitem{Kim:2005tw}
  Y.~Kim, B.~Kyae and J.~Lee,
  ``Global and local D-vortices,''
  JHEP {\bf 0510}, 002 (2005)
  [arXiv:hep-th/0508027];
  I.~Cho, Y.~Kim and B.~Kyae,
  ``DF-strings from D3 D3-bar as cosmic strings,''
  JHEP {\bf 0604}, 012 (2006)
  [arXiv:hep-th/0510218].


\bibitem{Gibbons:2000hf}
G.~W.~Gibbons, K.~Hori and P.~Yi,
``String fluid from unstable D-branes,''
Nucl.\ Phys.\ B {\bf 596}, 136 (2001)
[arXiv:hep-th/0009061].


\bibitem{Kwon:2003qn}
  O-K.~Kwon and P.~Yi,
  ``String fluid, tachyon matter, and domain walls,''
  JHEP {\bf 0309}, 003 (2003)
  [arXiv:hep-th/0305229].


\bibitem{Yee:2004ec}
  H.~U.~Yee and P.~Yi,
   ``Open / closed duality, unstable D-branes, and coarse-grained closed
  strings,''
  Nucl.\ Phys.\ B {\bf 686}, 31 (2004)
  [arXiv:hep-th/0402027].


\bibitem{Yi:1999hd}
P.~Yi, ``Membranes from five-branes and fundamental strings from
Dp branes,'' Nucl.\ Phys.\ B {\bf 550}, 214 (1999)
[arXiv:hep-th/9901159].


\bibitem{Bergman:2000xf}
O.~Bergman, K.~Hori and P.~Yi, ``Confinement on the brane,''
Nucl.\ Phys.\ B {\bf 580}, 289 (2000) [arXiv:hep-th/0002223].


\bibitem{Sen:1998tt}
  A.~Sen,
  ``SO(32) spinors of type I and other solitons on brane-antibrane pair,''
  JHEP {\bf 9809}, 023 (1998)
  [arXiv:hep-th/9808141].


\bibitem{Sen:1998ex}
  A.~Sen,
  ``BPS D-branes on non-supersymmetric cycles,''
  JHEP {\bf 9812}, 021 (1998)
  [arXiv:hep-th/9812031].

\bibitem{Sen:2003zf}
  A.~Sen,
  ``Moduli space of unstable D-branes on a circle of critical radius,''
  JHEP {\bf 0403}, 070 (2004)
  [arXiv:hep-th/0312003].




\bibitem{Garousi:2000tr}
M.~R.~Garousi, ``Tachyon couplings on non-BPS D-branes and
Dirac-Born-Infeld action,'' Nucl.\ Phys.\ B {\bf 584}, 284 (2000)
[arXiv:hep-th/0003122].


\bibitem{Bergshoeff:2000dq}
E.~A.~Bergshoeff, M.~de Roo, T.~C.~de Wit, E.~Eyras and S.~Panda,
``T-duality and actions for non-BPS D-branes,'' JHEP {\bf 0005},
009 (2000) [arXiv:hep-th/0003221].


\bibitem{Kluson:2000iy}
J.~Kluson, ``Proposal for non-BPS D-brane action,'' Phys.\ Rev.\ D
{\bf 62}, 126003 (2000) [arXiv:hep-th/0004106].


\bibitem{Sen:2003xs}
  A.~Sen,
  ``Open-closed duality at tree level,''
  Phys.\ Rev.\ Lett.\  {\bf 91}, 181601 (2003)
  [arXiv:hep-th/0306137].


\bibitem{Kim:2005pz}
  Y.~Kim and O-K.~Kwon,
  ``Exact rolling tachyon in noncommutative field theory,''
  Mod.\ Phys.\ Lett.\ A {\bf 21}, 421 (2006)
  [arXiv:hep-th/0501016].


\end{thebibliography}
\end{document}